\documentclass[conference]{IEEEtran}
\IEEEoverridecommandlockouts
\usepackage{cite}
\usepackage{comment}
\usepackage{algorithm}
\usepackage{algpseudocode}
\usepackage{amsmath,amssymb,amsfonts}
\usepackage{graphicx}
\usepackage[caption=false,font=footnotesize]{subfig}
\usepackage{textcomp}
\usepackage{xcolor}
\def\BibTeX{{\rm B\kern-.05em{\sc i\kern-.025em b}\kern-.08em
    T\kern-.1667em\lower.7ex\hbox{E}\kern-.125emX}}
\begin{document}

\title{Neural Multipath Fields for 6G FR3 Digital Twins: Latent Adaptation from Sparse Data\\

\thanks{This work was supported in part by the German Federal Ministry of
Research, Technology and Space (BMFTR) in the course of the 6GEM+ Transfer Hub
under grant 16KIS2411 and in part by the Deutsche Forschungsgemeinschaft (DFG, German Research Foundation) under Project–ID 287022738 TRR 196 for Project S02.}
}

\author{
    \IEEEauthorblockN{Srijita Sanyal, Adam Umra, Aydin Sezgin}
    \IEEEauthorblockA{Institute of Digital Communication Systems, Ruhr University Bochum, Germany
    \\\{srijita.sanyal, adam.umra, aydin.sezgin\}@rub.de}
}

\maketitle

\begin{abstract}

Wireless digital twins are essential for environment-aware 6G communications, motivating the investigation of frequency range 3 (FR3), an emerging mid-band regime offering a promising balance between bandwidth and propagation characteristics. Neural Radio Frequency (RF) fields provide continuous surrogates for site-specific radio propagation, but adapting a scene-specific field to a new environment can require substantial measurements and retraining. To this end, we propose the Latent-Adaptive Neural Multipath Field (LANMF), a scene-conditioned FR3 model that predicts path-resolved delays, complex channel impulse response (CIR) coefficients, and arrival and departure directions from transmitter–receiver coordinates. LANMF learns shared propagation representations across source environments and adapts to unseen scenes by optimizing only a 16-dimensional scene latent. Evaluated on six ray-traced indoor scenes at 7.5 GHz, latent-only adaptation yields improvements in prediction with just 5–10 target links (0.42–0.83\% of available links). Compared with full-network fine-tuning, it updates only 16 persistent target parameters, reducing adaptation time by 28–30\% and incremental peak GPU memory by 82–85\%. While the proposed solution supports various wireless applications, we demonstrate its utility through beam-candidate reduction using a 64-beam codebook, retaining measurement-based final beam selection.
\end{abstract}

\begin{IEEEkeywords}
Digital twin, FR3, 
Neural Radiance Field,
integrated sensing and communication, 
RF-NeRF,
few-shot adaptation, beam management
\end{IEEEkeywords}

\section{Introduction}
\label{sec:introduction}

Future 6G networks are expected to integrate communication, localization, and sensing, motivating the investigation of frequency range 3 (FR3), often termed the golden band. FR3 offers a promising balance between bandwidth and propagation characteristics for integrated sensing and communications
(ISAC) applications~\cite{baduge2026fr3,liu2022isac}. In ISAC systems, directional beamforming is important for both maintaining communication links and controlling the spatial illumination of the environment for sensing~\cite{umra2026rlaided}. Consequently, efficient beam management benefits from knowledge of the underlying propagation paths and their directional characteristics. Effective beam management therefore requires site-specific channel knowledge that captures how signals propagate through the surrounding environment.

Wireless digital twins and channel knowledge maps (CKMs) provide complementary approaches to representing such information~\cite{srijita,zeng2024tutorial}. While digital twins can model propagation through environment geometry and ray tracing, CKMs enable spatial channel inference from available observations. However, obtaining accurate channel knowledge across an environment can require substantial environmental information or measurements, motivating measurement-efficient prediction.

Coordinate-based neural representations address this problem by learning a function that maps spatial coordinates to channel-related quantities. These continuous and differentiable representations enable channel prediction at arbitrary spatial coordinates and support gradient-based optimization, making them suitable for spatial channel modeling and adaptation. Once trained, the network can be queried at new transmitter (Tx)--receiver (Rx) locations to predict the corresponding channel information. Inspired by Neural Radiance Fields (NeRFs), originally developed for novel-view synthesis~\cite{mildenhall2020nerf}, NeRF$^2$ applied this idea to radio-frequency (RF) propagation for localization and multiple-input multiple-output (MIMO) channel prediction~\cite{zhao2023nerf2}. Frequency-Aware NeRF$^2$ subsequently extended the approach toward wideband prediction~\cite{zhaofreqaware}. Spatial coordinate encodings can further help these networks capture rapid channel variations~\cite{mildenhall2020nerf}.

A model trained for one environment, however, cannot be assumed to generalize to another. Recent approaches therefore incorporate additional observations and information about the propagation environment. GRaF combines neighboring channel observations with scene geometry and neural ray tracing to predict spatial spectra~\cite{yang2026graf}. Voxel-CKM uses compact representations based on voxels, the three-dimensional counterparts of pixels, together with information about the transmitter for few-shot CKM reconstruction~\cite{li2026voxelckm}.

The methods discussed above primarily target RF radiance, spatial spectra, or channel state information, rather than jointly predicting individual multipath component (MPC) delays, complex coefficients, and arrival and departure directions. A path-resolved representation instead describes each propagation path through its delay, complex coefficient, and arrival and departure directions. These quantities reveal the temporal and directional structure of the channel and can support applications such as beam selection. However, conventional codebook-based beam selection may require probing a large number of candidate beams to identify the strongest communication link, introducing measurement overhead~\cite{liu2022isac}. Predicting the underlying multipath characteristics can help prioritize promising beams and reduce the number of candidates requiring measurement, in FR3 wireless systems. However, explicitly predicting MPCs in unseen environments from only a few observed links remains challenging. Full-network fine-tuning offers adaptation flexibility but increases
computational cost and the risk of overfitting to sparse observations \cite{basu2023strongbaselinesparameterefficient}. In contrast, latent-only
adaptation updates a compact scene representation while preserving shared model parameters, trading adaptation flexibility for parameter efficiency~\cite{park2019deepsdf}.

Motivated by this challenge, we propose the \emph{Latent-Adaptive Neural Multipath Field (LANMF)} for predicting path-resolved channel parameters from Tx–Rx coordinates. LANMF learns shared network parameters across source environments alongside compact scene-specific latent representations. For an unseen environment, only the scene latent is optimized using a few observed links, while the pretrained network remains fixed. We evaluate prediction accuracy and adaptation cost under limited target observations through comparisons with interpolation, target-only learning, and full-network fine-tuning. Beyond channel prediction, the proposed path-resolved representation can support various wireless applications, including localization, sensing, and beam management. As an initial application within the broader context of ISAC-enabled wireless systems, we investigate beam-candidate reduction in communication. Predicted multipath information identifies promising transmit beams, reducing the number of candidates requiring channel probing to determine the final communication beam.

% ============================================================
\section{System Model}
\label{sec:system_model}
% ============================================================

To formalize the problem, we consider a collection of indoor propagation environments indexed
by $s$. Each environment represents a bounded indoor space characterized by its geometry, reflecting surfaces, and electromagnetic material properties, determining multipath propagation conditions within that space. Communication takes place at a constant FR3 regime. The transmitters use an $8\times8$ planar antenna array, while the receivers use a single antenna.
Each wireless link observed in each environment is defined by the transmitter and receiver
positions
$\mathbf{p}_{\mathrm{Tx}},
\mathbf{p}_{\mathrm{Rx}}\in\mathbb{R}^{3}$.
The corresponding coordinate input is
\begin{equation}
\mathbf{x}
=
\begin{bmatrix}
\mathbf{p}_{\mathrm{Tx}} \\
\mathbf{p}_{\mathrm{Rx}}
\end{bmatrix}
\in\mathbb{R}^{6}.
\label{eq:coordinate_input}
\end{equation}

For  $P$ resolvable MPCs, the baseband channel impulse response (CIR) is represented as
\begin{equation}
h_s(t;\mathbf{x})
=
\sum_{p=1}^{P}
m_{s,p}\,h_{s,p}\,
\delta\!\left(t-\tau_{s,p}\right),
\label{eq:cir_model}
\end{equation}
where $\tau_{s,p}$ and $h_{s,p}\in\mathbb{C}$ denote the propagation delay and
complex channel coefficient of path $p$, respectively, $\delta(\cdot)$ is the Dirac delta
distribution representing an impulse at the corresponding path delay. To ensure a consistent input size for the neural networks, each Tx–Rx link is represented by $P$ path entries, with unused entries padded when fewer than $P$ resolvable MPCs are present. The binary mask $m_{s,p}$ identifies valid paths: it equals 1 for a resolvable MPC and 0 for a padded entry.
Each valid MPC is represented by
\begin{equation}
\mathcal{M}_{s,p}
=
\left\{
\tau_{s,p},
h_{s,p},
\mathbf{u}^{\mathrm{AoA}}_{s,p},
\mathbf{u}^{\mathrm{AoD}}_{s,p}
\right\},
\label{eq:mpc_tuple}
\end{equation}
where $\mathbf{u}^{\mathrm{AoA}}_{s,p},
\mathbf{u}^{\mathrm{AoD}}_{s,p}\in\mathbb{S}^{2}$ are unit-vector
representations of the arrival and departure directions
with $\mathbb{S}^{2}=
\{\mathbf{u}\in\mathbb{R}^{3}:
\|\mathbf{u}\|_2=1\}$ representing the unit sphere.
The phase is derived from the complex path coefficient rather than predicted directly,
\begin{equation}
\phi_{s,p}=\arg(h_{s,p}).
\label{eq:phase_definition}
\end{equation}

The resulting MPC representation characterizes the propagation channel for each Tx--Rx link. The proposed framework addresses two interrelated challenges, namely, representing environment-dependent multipath propagation and adapting channel predictions to unseen environments with limited observations. In addition to prediction accuracy, the adaptation process should minimize the number of target-specific parameters requiring optimization, as well as the associated computational and memory overhead.

% ============================================================
\section{Latent-Adaptive Neural Multipath Field}
\label{sec:experimental}
% ============================================================

To address environment-dependent multipath propagation and sparse adaptation to unseen environments, we propose LANMF. The framework comprises two stages: learning a shared scene-conditioned multipath representation across source environments and adapting a compact latent embedding to an unseen environment.

%%%%%%%%%%%%%%%%%%%%%%%%%%%%%%%%%%%%%%%%%%%%%%%%%
%%%%%%%%%%%%%%%%%% New Figure %%%%%%%%%%%%%%%%%%
\begin{figure*}[htpb]
    \centering
    \subfloat[Source pretraining.
        \label{fig:lanmf-pretraining}]{%
        \includegraphics[width=0.47\textwidth]
            {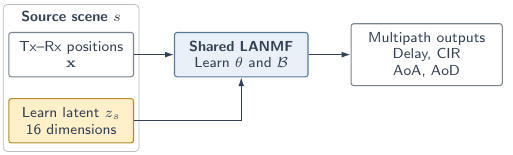}%
    }%
    \hspace{0.06\textwidth}%
    \subfloat[Few-shot target adaptation.
        \label{fig:lanmf-adaptation}]{%
        \includegraphics[width=0.47\textwidth]
            {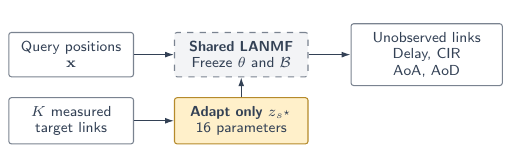}%
    }

    \caption{LANMF overview. (a) Jointly learn the shared field
    $(\theta,B)$ and source-scene latents. (b) Freeze the shared
    field and update only a 16-dimensional target latent using
    $K$ measured links.}
    \label{fig:lanmf_framework}
\end{figure*}
%%%%%%%%%%%%%%%%%%%%%%%%%%%%%%%%%%%%%%%%%%%%%%%%%
%%%%%%%%%%%%%%%%%%%%%%%%%%%%%%%%%%%%%%%%%%%%%%%%%

\subsection{Shared Multipath Field Learning}
\label{sec:lanmf_architecture}

 The source-learning workflow is illustrated in Fig.~\ref{fig:lanmf-pretraining}. Conventional coordinate-based neural fields map coordinates to channel characteristics but do not explicitly account for environmental variations in geometry, materials, and scattering conditions. LANMF addresses this dependence by conditioning a shared neural field on a scene-specific latent representation:
\begin{equation}
f_{\boldsymbol{\theta}}:
(\mathbf{x},\mathbf{z}_s)
\longmapsto
\left\{
\widehat{\mathcal{M}}_{s,p}
\right\}_{p=1}^{P},
\label{eq:field_mapping}
\end{equation}
where $\widehat{\mathcal{M}}_{s,p}$ denotes the predicted parameters of MPC $p$ in environment $s$, including propagation delay, complex CIR coefficient, AoA, and AoD. The parameters $\boldsymbol{\theta}$ are shared across environments, whereas $\mathbf{z}_s$ captures scene-specific propagation variations.

To represent fine-scale spatial variations, the normalized Tx--Rx coordinates defined
in (\ref{eq:coordinate_input}) are mapped using sinusoidal positional encoding
~\cite{mildenhall2020nerf}. For each scalar coordinate $q$,
\begin{equation}
\gamma(q)
=
\left[
q,
\left\{
\sin(2^{\ell}\pi q),
\cos(2^{\ell}\pi q)
\right\}_{\ell=0}^{L-1}
\right],
\label{eq:positional_encoding}
\end{equation}
where $L$ is the number of frequency bands.The encoding retains each coordinate and adds $2L$ sinusoidal features, producing a $6(2L+1)$-dimensional multiscale representation for the six input coordinates.

Each source environment is assigned a trainable latent embedding
$\mathbf{z}_s\in\mathbb{R}^{d_z}$, which is concatenated with the encoded coordinates to form
\begin{equation}
\mathbf{v}_s =
[\gamma(\mathbf{x});\mathbf{z}_s]
\in\mathbb{R}^{6(2L+1)+d_z}.
\label{eq:lanmf_input}
\end{equation}

The resulting representation is processed by a shared multilayer perceptron (MLP) with separate channel and directional output heads. For each path slot $p$, the channel head predicts
\begin{equation}
\widehat{\mathbf{c}}_p=
\left[
\widehat{\tau}_p,
\Re\{\widehat h_p\},
\Im\{\widehat h_p\}
\right],
\label{eq:channel_head}
\end{equation}
where $\widehat{\tau}_p$ is the propagation delay and $\widehat h_p$ is the complex CIR coefficient. Delays are standardized during training for numerical stability and converted back to physical units during inference, while phase is obtained from (\ref{eq:phase_definition}).

Rather than predicting only aggregate channel responses, the proposed model explicitly reconstructs the individual multipath characteristics defined in (\ref{eq:mpc_tuple}), preserving the temporal and directional information required for prioritizing promising transmit beams before measurement.

In parallel, the directional heads predict angular coefficients over separate learnable AoA and AoD bases. For $\kappa\in\{\mathrm{AoA},\mathrm{AoD}\}$,
\begin{equation}
\widehat{\mathbf{u}}^\kappa_p=
\frac{
\sum_{j=1}^{J}
\alpha^\kappa_{p,j}\mathbf{b}^\kappa_j
}{
\left\|
\sum_{j=1}^{J}
\alpha^\kappa_{p,j}\mathbf{b}^\kappa_j
\right\|_2+\epsilon
},
\qquad J=16,
\label{eq:direction_basis}
\end{equation}
where $\alpha^\kappa_{p,j}$ denotes the predicted angular coefficient and $\mathbf{B}^{\kappa} =[\mathbf{b}_1^{\kappa},\ldots, \mathbf{b}_J^{\kappa}]^T \in\mathbb{R}^{J\times3}$ contains the learnable directional bases.
The complete basis collection is $\mathcal{B}= \{\mathbf{B}^{\mathrm{AoA}}, \mathbf{B}^{\mathrm{AoD}}\}$.

Shared directional bases encourage transferable angular representations, while scene- and link-dependent coefficients capture directional variations. Normalized three-dimensional AoA/AoD vectors avoid angular-coordinate discontinuities, and $J$ balances representational flexibility and complexity. The basis size $J$ is selected to balance representational flexibility and model complexity.

The shared network, directional bases, and source-scene embeddings are jointly optimized using valid MPC observations. The source-training objective is
\begin{equation}
\mathcal{L}_{\mathrm{src}}
=
5\mathcal{L}_{\mathrm{ch}}
+
\sum_{\kappa\in\{\mathrm{AoA},\mathrm{AoD}\}}
\left(
2\mathcal{L}^{\kappa}_{\mathrm{circ}}
+
\mathcal{L}^{\kappa}_{\mathrm{dir}}
\right)
+
10^{-3}\mathcal{L}_{\mathrm{basis}},
\label{eq:training_loss}
\end{equation}
where $\mathcal{L}_{\mathrm{ch}}$ is the masked reconstruction loss over standardized delays and complex CIR components, $\mathcal{L}_{\mathrm{circ}}$ accounts for angular periodicity, and
\begin{equation}
\mathcal{L}^{\kappa}_{\mathrm{dir}}
=
\frac{1}{\sum_p m_p}
\sum_{p}m_p
\left(
1-
\widehat{\mathbf{u}}^{\kappa T}_{p}
\mathbf{u}^{\kappa}_{p}
\right),
\label{eq:direction_loss}
\end{equation}
penalizes directional disagreement over valid paths, indicated by $m_p$. The regularizer $\mathcal{L}_{\mathrm{basis}}$ discourages collapse of the learned directional primitives. Joint source training therefore learns a common multipath representation while allowing individual scenes to exhibit different propagation characteristics through their latent embeddings.

\subsection{Few-Shot Latent Adaptation}
\label{sec:latent_adaptation}

Unseen environments may differ from source scenes in geometry, materials, and scattering conditions. Conventional full-network fine-tuning updates large number of pretrained parameters, increasing adaptation cost and potentially exacerbating overfitting under limited target observations~\cite{basu2023strongbaselinesparameterefficient}.

To address this second challenge, LANMF extends the shared-field representation through latent-only adaptation (LANMF-LA). For an unseen environment $s^\star$, the pretrained network $\boldsymbol{\theta}^\star$ and directional bases $\mathcal{B}^\star$ are frozen, while a new scene latent $\mathbf{z}_{s^\star}\in\mathbb{R}^{d_z}$ is initialized and optimized using the available target measurements.  The adaptation workflow is illustrated in Fig.~\ref{fig:lanmf-adaptation}. At each adaptation step, the support-link coordinates and current scene latent are used to predict MPC parameters, which are compared with the observed MPCs through a masked reconstruction loss. The resulting gradients iteratively update only the scene latent, allowing it to capture the propagation characteristics of the unseen environment.

Given a support set $\mathcal{D}_{\mathrm{sup}}^K$ containing $K$ observed Tx--Rx links, the target latent is estimated as
\begin{equation}
\mathbf{z}_{s^\star}^{\star}
=
\arg\min_{\mathbf{z}_{s^\star}}
\frac{1}{K}
\sum_{i=1}^{K}
\mathcal{L}_{\mathrm{MPC}}
\left(
f_{\boldsymbol{\theta}^\star}
(\mathbf{x}_i,\mathbf{z}_{s^\star};
\mathcal{B}^\star),
\mathcal{Y}_i
\right),
\label{eq:latent_adaptation}
\end{equation}
where $\mathcal{L}_{\mathrm{MPC}}$ denotes the masked multipath reconstruction objective and $\mathcal{Y}_i$ contains the observed MPC parameters and validity indicators for support link $i$.

By updating only $d_z$ target-specific latent parameters, LANMF preserves the shared field and limits the number of free parameters estimated from sparse observations. Restricting optimization to the scene latent therefore reduces the number of trainable target-specific parameters and the associated optimizer-state GPU requirements when compared with full-network fine-tuning.

Once adapted, the optimized latent is reused to predict multipath characteristics at previously unobserved target links:
\begin{equation}
\widehat{\mathcal{Y}}(\mathbf{x})
=
f_{\boldsymbol{\theta}^\star}
(\mathbf{x},\mathbf{z}_{s^\star}^{\star};
\mathcal{B}^\star),
\quad
\mathbf{x}\in\mathcal{D}_{\mathrm{qry}}^{x},
\label{eq:target_prediction}
\end{equation}
where $\mathcal{D}_{\mathrm{qry}}^{x}$ contains the coordinates of target query links disjoint from the adaptation support set. Thus, the model transfers a pretrained multipath field to an unseen environment by adapting its scene representation rather than retraining the network. This enables channel prediction across unobserved Tx--Rx configurations using limited target measurements. 

Beyond channel reconstruction, the predicted MPCs support transmit-beam prioritization. We consider a transmitter equipped with an antenna array and a predefined beamforming codebook $\mathcal{B}_{\mathrm{beam}}$, where each beam $b\in\mathcal{B}_{\mathrm{beam}}$ corresponds to a beamforming vector $\mathbf{w}_b$. For each candidate beam, the predicted MPC delays, complex coefficients, and angles of departure (AoDs) are combined with the array response to estimate its received power $\widehat{P}_b$. This enables the beams to be ranked without first measuring every candidate. The $N$ highest-ranked beams form the candidate set 
\begin{equation}
\mathcal{C}_{N}
=
\operatorname{TopN}_{b\in\mathcal{B}_{\mathrm{beam}}}
\widehat{P}_{b},
\label{eq:candidate_set}
\end{equation}
which can be used to restrict subsequent beam probing to the most promising candidates. The final beam is selected from these measured candidates, reducing the beam-search space while aiming to retain an optimal or near-optimal beam.

\subsection{Alternative strategy: Full Fine Tuning}

To assess the effect of restricting adaptation to the scene latent, we also consider full-network fine-tuning (LANMF-FT). LANMF-FT adapts the pretrained network parameters using target support observations instead of only the scene latent, isolating the effect of the adaptation strategy under the same architecture.

\section{Results and Discussion}

Simulations use channels generated using the Sionna Ray Tracer~\cite{hoydis2023sionna} for six digital twins of indoor environments at a carrier frequency of 7.5~GHz. The digital twins were created using Blender 4.2. Each scene contains three different transmitters positions and for each, 400 different receiver locations are sampled, yielding 1,200 Tx--Rx links per scene and 7200 data points in total. The maximum number of path slots is $P=32$. Each observation contains Tx/Rx coordinates, path delays, complex CIR coefficients, AoA, AoD, validity masks, path count, and LoS labels. The masks are used to exclude invalid slots from supervised losses.

Digital Twin channels of environments 0--3 constitute the source-training pool. Environment~4 is retained as a development environment for model selection, operating-point selection, and adaptation choices, while Scene~5 is used as the unseen few-shot target. Additional architectural and neural network parameters are listed in Table~\ref{tab:lanmf_parameters}.
\begin{table}[!htbp]
\centering
\caption{LANMF Architecture and Training Configuration}
\label{tab:lanmf_parameters}
\footnotesize
\renewcommand{\arraystretch}{1.15}
\begin{tabular}{ll}
\hline
\textbf{Parameter} & \textbf{Value} \\
\hline
\multicolumn{2}{l}{\textit{Network Architecture}} \\
\hline
Positional encoding frequencies ($L$) & 6 \\
Scene latent dimension ($d_z$) & 16 \\
MLP Layers  & 3 \\
Neurons per layer  & 256 \\
Directional basis vector ($J$)  & 16 \\
\hline
\multicolumn{2}{l}{\textit{Source Training}} \\
\hline
Optimizer & AdamW \\
Learning rate & $3\times10^{-4}$ \\
Weight decay & $10^{-5}$ \\
Batch size & 64 \\
Training epochs & 50 \\
Gradient clipping & 1.0 \\
\hline
\multicolumn{2}{l}{\textit{LANMF-LA Adaptation}} \\
\hline
Optimizer & AdamW \\
Learning rate & $5\times10^{-2}$ \\
Adaptation steps & 200 \\
Adaptation batch size & $\leq64$ \\
Gradient clipping & 1.0 \\
Trainable target parameters & 16 \\
Support links ($K$) & 1, 2, 5, 10, 20, 50, 100 \\
Support-set seeds & 5 \\
\hline
\end{tabular}
\end{table}

Our proposed model is compared with baselines and ablations to assess source transfer, scene conditioning, and latent-only adaptation. For ease of comprehension, in the results, our proposed latent-only adaptation method is denoted as LANMF-LA and the full fine tuned model as LANMF-FT, which acts as an ablation denoting the effects of our adaptation strategy. Another baseline NeRF-FT uses a conventional coordinate-based neural field~\cite{diff_rt_calibration_2023} to predict multipath parameters directly, without scene latent or learned directional bases. KNN provides nearest-neighbor interpolation following GRaF~\cite{yang2026graf}, as another baseline. Target-NF acts as another important baseline, training a randomly initialized neural field solely on target support links, without source pretraining, to assess the benefits of transferable multipath representations.

Adaptation efficiency is evaluated separately through \emph{(i) persistently changed target-specific parameters, (ii) optimizer-tracked parameters and optimizer-state storage, (iii) wall-clock adaptation time}, and \emph{(iv) incremental peak GPU memory.}

\subsection{Comparison with Baselines and Adaptation Ablations}

\label{sec:results}
\begin{figure*}[htpb]
        \centering
        \includegraphics[width=1\linewidth]{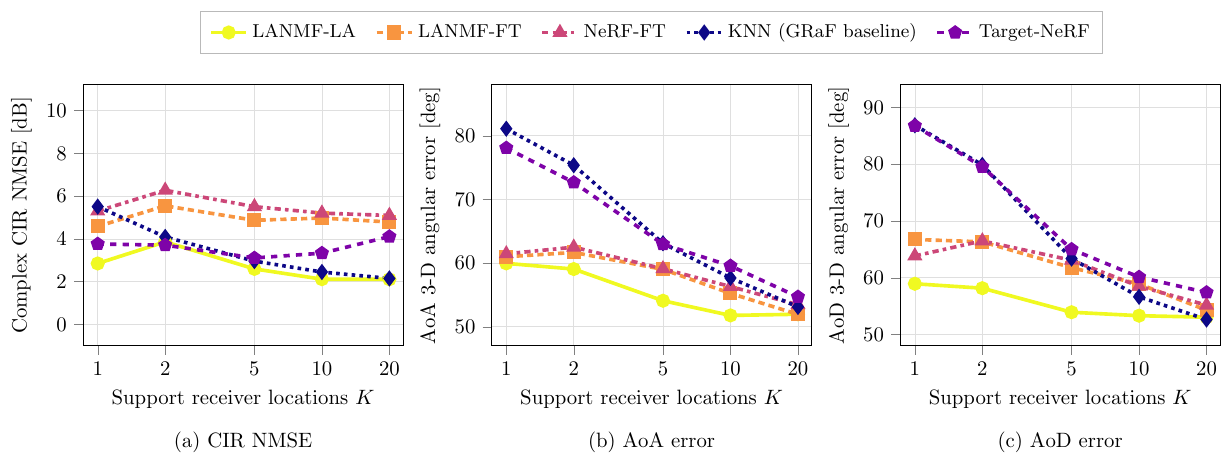}
        \caption{Few-shot multipath prediction on an unseen target scene. (a) Complex CIR NMSE, (b) AoA 3-D angular error, and (c) AoD 3-D angular error versus the number (K) of support Tx–Rx channel observations.}
        \label{fig:placeholder}
\end{figure*}
\begin{table}[t]
\centering
\caption{LANMF-LA Multipath Prediction Errors on the Unseen Target Scene}
\label{tab:lanmf_results}
\scriptsize
\setlength{\tabcolsep}{2pt}
\renewcommand{\arraystretch}{1.1}

\begin{tabular}{c|c|c|c|c}
\hline
$K$ & Delay [ns] & CIR [dB] & AoA [$^\circ$] & AoD [$^\circ$] \\
\hline
1   & $8.89 \pm 1.76$ & $2.85 \pm 1.39$ & $59.95 \pm 6.54$ & $58.94 \pm 3.53$ \\
5   & $6.90 \pm 0.51$ & $2.60 \pm 0.89$ & $54.08 \pm 1.55$ & $53.92 \pm 1.24$ \\
10  & $6.31 \pm 0.46$ & $2.12 \pm 0.48$ & $51.77 \pm 0.51$ & $53.31 \pm 0.50$ \\
20  & $6.35 \pm 0.46$ & $2.11 \pm 0.37$ & $51.96 \pm 0.57$ & $53.04 \pm 0.61$ \\
100 & $6.17 \pm 0.14$ & $1.87 \pm 0.19$ & $51.46 \pm 0.31$ & $52.71 \pm 0.45$ \\
\hline
\end{tabular}

\vspace{1mm}
\parbox{0.95\columnwidth}{\centering\scriptsize
Delay, AoA, and AoD are MAE; CIR is NMSE.
Values are mean $\pm$ standard deviation over five seeds.
}
\end{table}
Table \ref{tab:lanmf_results} summarizes the prediction performance of LANMF-LA for different numbers of target observations. The results indicate that the main benefit of LANMF-LA appears in the strongly data-limited regime. Comparisons with the baselines are presented in Fig. \ref{fig:placeholder}. AoA and AoD errors are evaluated using the three dimensional geodesic direction error~\cite{hannan2025doa}. For both AoA and AoD, LANMF-LA exhibits substantially lower errors than the alternatives for $K\leq20$, as shown in Fig. \ref{fig:placeholder}(b) and Fig. \ref{fig:placeholder}(c).At $K=1$, LANMF-LA remains near $60^\circ$ for both directional quantities, whereas KNN and Target-NF exhibit considerably larger errors and substantially larger support-set variability. As $K$ increases, the differences narrow, suggesting a favorable trade-off between adaptation capacity and the amount of available target data. The delay prediction error is calculated using masked mean absolute error~\cite{huang2022aiwireless}. Table \ref{tab:lanmf_results} shows that although the delay is decreasing with more observation points, it tanks after the mid-regime and there is room for improvement in terms of prediction.

 To assess the channel prediction error, we use the CIR NMSE~\cite{arvinte2022mimochannelestimationusing}. In the results as shown in Figure \ref{fig:placeholder}(a), LANMF-LA obtains the error among the learned models over much of the low-$K$ range and improves to approximately $2.1$~dB by $K=10$--$20$, shown in Table \ref{tab:lanmf_results}. In contrast, LANMF-FT and NeRF-FT remain near $5$~dB over this range. KNN becomes increasingly competitive as additional spatial samples are available, illustrating that local interpolation is a strong reference for aggregate channel reconstruction. 

\subsection{Parameter and Computational Efficiency}

The preceding results establish that restricting adaptation to the scene latent can improve robustness when target data are scarce. The second motivation for latent adaptation is deployment efficiency. The complete LANMF contains $443{,}936$ parameters, whereas LANMF-LA persistently modifies only the $16$ elements of the target-scene latent. The training and simulation configurations of our model are shown in Table \ref{tab:lanmf_parameters}.

Table~\ref{tab:efficiency} compares LANMF-LA with unrestricted LANMF-FT under the identical adaptation budget. Depending on $K$, LANMF-FT changes approximately $3.31$--$3.50\times10^5$ model elements, whereas LANMF-LA changes only $16$, corresponding to a reduction of more than $99.995\%$ in persistently adapted parameters. Optimizer-state storage decreases from approximately $3.387$~MB to $7.36\times10^{-4}$~MB.

\begin{table}[t]
\centering
\caption{Efficiency Improvement of LANMF-LA over LANMF-FT}
\label{tab:efficiency}
\scriptsize
\setlength{\tabcolsep}{5pt}

\begin{tabular}{c|c|c|c}
\hline
$K$ &
Adapted Params. $\downarrow$ &
Adaptation Time $\downarrow$ &
Peak GPU Memory $\downarrow$ \\
\hline
5  & 99.995\% & 28.2\% & 84.5\% \\
10 & 99.995\% & 29.9\% & 83.8\% \\
20 & 99.995\% & 30.1\% & 82.5\% \\
\hline
\end{tabular}

\vspace{1mm}
\parbox{0.98\columnwidth}{\scriptsize
Values indicate the percentage reduction achieved by LANMF-LA relative
to unrestricted LANMF-FT. GPU memory denotes incremental peak allocated
memory during target adaptation.
}
\end{table}

Consequently, the measured computational reduction also remains substantial: LANMF-LA reduces wall-clock adaptation time by $28.2$--$30.1\%$ and incremental peak allocated GPU memory by $82.5$--$84.5\%$ for $K=5$--$20$. The result therefore supports our claim that LANMF-LA is \emph{parameter-efficient and computationally lighter} than full fine-tuning, rather than implying a $99.995\%$ reduction in total training complexity.

\subsection{Neural-Field-Assisted Beam Candidate Reduction}
We evaluate the beam-management benefits of LANMF using a 64-beam DFT codebook for the $8\times8$
transmit array. 
\begin{figure}[htpb]
    \centering
    \includegraphics[width=0.95\linewidth]{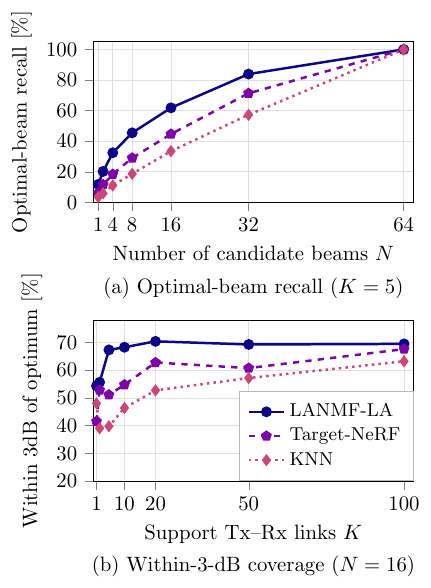}
    \caption{Beam candidate reduction:
    (a) optimal-beam recall versus retained candidates
    $N$ at $K=5$; (b) within-3-dB coverage versus
    support links $K$ at $N=16$. Final beam selection
    remains measurement-based.}
    \label{fig:beam_results}
\end{figure}
The predicted beam power used to construct $\mathcal{C}_N$ in (\ref{eq:candidate_set})
is computed as
\begin{equation}
\widehat P_b
=
\frac{1}{N_f}
\sum_{n=1}^{N_f}
\left|
\sum_{p\in\mathcal{P}}
\widehat h_p\,
\mathbf{w}_{b}^{H}
\mathbf{a}
\left(
\widehat{\mathbf{u}}^{\mathrm{AoD}}_{p}
\right)
e^{-j2\pi f_n\widehat{\tau}_p}
\right|^2,
\label{eq:wideband_beam_score}
\end{equation}
where $\mathbf{a}(\cdot)$ the array response, $f_n$ the frequency sample, $N_f$ the number of frequency samples, and $\mathcal{P}$ the considered MPCs.

Performance is assessed through optimal-beam recall (Top-$N$ accuracy), beam-selection power
loss relative to exhaustive search, and beam-search reduction~\cite{hashir2025gps}. We additionally report within-3-dB coverage, indicating the fraction of links retaining a candidate within 3 dB of the optimal beam.

As shown in Fig.~\ref{fig:beam_results}, LANMF-LA outperforms KNN and Target-NF in both optimal-beam recall and within-3-dB coverage, with most adaptation gains occurring by $K=5$--$20$. At $N=16$, only 25\% of beams require probing, reducing the search space by 75\%. With $K=5$, optimal-beam recall
improves from 44.2\% (zero-shot) to 61.5\%, while within-3-dB coverage increases from 50.3\% to 66.7\%. At $N=32$ and $K=5$, recall reaches 82.8\%, with 85.9\% within-3-dB coverage, using only 0.42\% of target links for adaptation.

These results demonstrate that LANMF can prioritize promising beams despite imperfect MPC reconstruction, reducing beam-search overhead while retaining measurement-based final beam selection.

\section{Conclusion}

This work addresses the challenge of predicting path-resolved wireless channels in unseen FR3
environments using limited observations while maintaining computationally efficient adaptation.
The results demonstrate that LANMF achieves measurement-efficient adaptation through compact scene representations and supports practical applications such as beam-candidate reduction. Future work will explore the integration of LANMF into broader ISAC applications, including environment-aware beam management and localization.

\bibliographystyle{IEEEtran}
\bibliography{bibliography}

\end{document}